# Orientation-dependent deformation mechanisms of bcc niobium nanoparticles

J.J. Bian[a,b], L. Yang[a], X.R. Niu[b], and G.F. Wang[a,*]

[a] Department of Engineering Mechanics, SVL, Xi'an Jiaotong University, Xi'an 710049, PR China

[b] CASM and Department of Mechanical and Biomedical Engineering, City University of Hong Kong, Hong Kong SAR, PR China

**Abstract**

Nanoparticles usually exhibit pronounced anisotropic properties, and a close insight into the atomic-scale deformation mechanisms is of great interest. In present study, atomic simulations are conducted to analyze the compression of bcc nanoparticles, and orientation-dependent features are addressed. It is revealed that surface morphology under indenter predominantly governs the initial elastic response. The loading curve follows the flat punch contact model in [110] compression, while it obeys the Hertzian contact model in [111] and [001] compressions. In plastic deformation regime, full dislocation gliding is dominated in [110] compression, while deformation twinning is prominent in [111] compression, and these two mechanisms coexist in [001] compression. Such deformation mechanisms are distinct from those in bulk crystals under nanoindentation and nanopillars under compression, and the major differences are also illuminated. Our results provide an atomic perspective on the mechanical behaviors of bcc nanoparticles and are helpful for the design of nanoparticle-based components and systems.

* Corresponding author. *E-mail address*: wanggf@mail.xjtu.edu.cn.

# 1. Introduction

Nanoparticles have attracted much attention owing to outstanding physical, mechanical and chemical properties, compared to their macroscopic counterparts.[1] Recently, rapid advances in synthesis technology have facilitated the fabrication of nanoparticles [2] and laid the foundations for their applications in such fields as high performance catalysts, antimicrobial materials and energy harvesting.[3] To maximize the potential of these applications, it is of critical importance to understand the mechanical properties of nanoparticles and unravel the deformation mechanisms, and many efforts have been devoted to this subject.

The pioneering researches in this field focus mainly on silicon nanoparticles, and rich mechanical phenomena and deformation mechanisms have been revealed. Gerberich et al. measured the hardness of spherical silicon nanoparticles, which is about four times as that of bulk silicon.[4] Under repeated compression, a cumulative reversible strain was observed up to ~ 0.4.[5] Moreover, it was found that the fracture toughness of silicon nanoparticles greatly increases compared to bulk silicon.[6] Molecular static studies showed that phase transition is the dominated deformation mechanism inside ultra-small silicon nanoparticles.[7] Further molecular dynamics (MD) researches demonstrated that dislocations can also nucleate and glide inside the phase-transition region, and the deformation mechanisms of silicon nanoparticle are size-dependent.[8] Owing to different constraint states in bulk materials and nanoparticles, it was concluded that silicon nanoparticles mainly display dislocation driven plasticity.[9] For hollow silicon nanoparticles, Yang et al. found that wall thickness affects both dislocation nucleation and failure modes.[10]

Analogous to silicon nanoparticles, metallic nanoparticles also exhibit improved mechanical properties and unique behaviors compared with conventional materials. MD simulations showed that the elastic moduli and yielding stresses of golden nanoparticles are size dependent.[11][12] Experiments confirmed that golden nanoparticles possess higher modulus and yielding stress than those of bulk gold.[13] Using *in-situ* high-resolution transmission electron microscopy, it was found that sub-10-nm silver nanoparticle can even deform like a liquid droplet at room temperature.[14] Also, reversible dislocation plasticity was observed inside some ultra-small silver nanoparticles.[15] Based on MD simulations, Dan Mordehai et al. found the pseudoelastic behaviors in the adhesive contact between two polyhedral nanoparticles.[16] In addition, the introduction of twin boundaries greatly enhances the strength and malleability of metallic nanoparticles.[17][18] Recently, Yang et al reported that the atomic-scale surface morphology induces the fluctuation of yield stress of golden nanoparticles.[19] Plasticity of nanoparticles usually initiates with nucleation of partial dislocation from contact fridge and characteristic pyramid hillocks.[20][21] The existing researches mainly consider the nanoparticles with face-centered-cubic (fcc) lattice structure, and reveal their distinct mechanical properties and behaviors.

This work focuses on body-centered-cubic (bcc) single crystalline metallic nanoparticles, which has been seldom addressed and is lack of consensus as to the deformation mechanism. Different from dislocations in fcc crystals, dislocations in bcc crystals are usually involved multiple slip systems [22] and have non-planar core structures.[23]-[25] Owing to a high surface-to-volume ratio of nanoparticles, the interplay between internal defects and free surface makes the deformation in bcc nanoparticle even more complicated. In present study, we perform a series of MD simulations to investigate the compression of bcc niobium (Nb) nanoparticles.

**2. Simulation details**

Fig. 1 presents the compression of a spherical bcc Nb nanoparticle, together with its typical initial configuration. Spherical nanoparticles are carved out of perfect single crystalline bulk Nb with a lattice constant of 3.30 Å. A nanoparticle with radius of 10 nm containing about 0.23 million atoms is considered. We utilize an embedded atom method (EAM) potential for single crystal Nb, which is fitted based on the force matched method.[26]

To simulate the uniaxial compression of a nanoparticle by two rigid planes, a repulsive potential is adopted to describe the interactions between the rigid planes and Nb atoms, which is given as

$$U_i(z_i) = \begin{cases} K(z_i - h_T)^3 & z_i \geq h_T \\ 0 & h_b < z_i < h_T \\ K(h_B - z_i)^3 & z_i \leq h_B \end{cases}, \quad (1)$$

where $K$ is a specified constant representing the rigidity of the planar indenter. Compression direction is parallel to the $z$-axis, and $z_i$, $h_T$ and $h_B$ represent the positions of the $i$-th atom, the top indenter, and the bottom indenter, respectively.

The open-source molecular dynamics simulator LAMMPS is adopted. All simulations are performed at constant temperature 10 K, and canonical (NVT) ensemble is utilized to describe the atomic system. The Nose-Hoover thermostat is used to control the temperature of the atomic systems.[27][28] Time integration is implemented based on the velocity-Verlet algorithm with a time step of 2.0 fs. After the construction of nanoparticle, the initial structure is firstly relaxed using the conjugate gradient method to reach a local stable state. Then the whole system is equilibrated at 10 K for about 20 ps. After well equilibration, the top and bottom indenters are moved simultaneously towards the center of the nanoparticle at a speed of 0.1 Å/ps. During compression process, atoms adjacent to indenters will experience repulsive forces. We denote the compressive strain by the ratio of compression depth $\delta$ of one indenter to the radius of nanoparticle $R$. In loading procedure, a total strain of ~ 30 % is applied to the nanoparticle. Delaunay triangulation algorithm is utilized to compute the contact area, and the averaged contact stress is defined as the compressive load divided by the current contact area.

In order to characterize the evolution of atomic defects inside the nanoparticles during yielding and post-yielding deformation stages, local atom structures are specified by the common neighbor analysis (CNA) parameter.[29] According to this method, atoms in defects such as surfaces, grain boundaries, twin boundaries and dislocation cores are distinguished from those in perfect lattice, and the details of dislocations could be extracted from the deformed structures. Slip traces of dislocations can be specified by slip vector.[30] In simulations, all atomic configurations are visualized by OVITO.[31] And dislocation extraction algorithm (DXA) is applied to extract dislocation lines from the deformed bcc nanoparticles. [32]

3. **Simulation results and discussions**

In different loading directions, owing to the distinction in surface morphology and optional activated slip systems in plasticity, the overall compressive response and underlying atomic deformation mechanism might be different. Therefore, we consider the compressions under three typical orientations including [110], [111] and [001].

**3.1 [110] compression**

Fig. 2 displays the loading responses of nanoparticles under [110] compression. In panel (a), the compressive load increases linearly up to the yielding point where $\delta$ is ~ 4.0 Å (it is confirmed by checking the nucleation of initial dislocations). After the onset of plasticity, load fluctuates with further compression progressing. Panel (b) shows the variations of contact area and averaged contact stress. In elastic stage, contact area keeps nearly constant and contact stress increases linearly with strain. In incipient plastic deformation after yielding, contact area exhibits a step-plateau behavior, and contact stress increases in serrated manner. When the compressive strain is larger than 0.10, the step-plateau behavior of contact area is no longer obvious, meanwhile contact stress keeps at a low level, and the magnitude of its fluctuation also decreases.

The underlying atomic originations are rendered as follows. In bcc crystal, the family of {110} crystalline planes are most close-packed, and {110} surface facets could be obviously observed in Fig. 1. When compression is conducted along [110] direction, indenter firstly touches the top-most atom layer. In elastic regime, the contact area equals to the area of the top-most [110] surface facet, and both the compressive load and contact stress linearly increase with respect to indent depth. When the atomic step is totally flattened, contact area abruptly increases, and dislocations nucleate around its fridges. Emergence of initial dislocation and increase of contact area lead to the large drop of contact stress.

Surface steps dominate the elastic response of nanoparticles. Due to the discrete surface steps on nanoparticles, Luan and Robbins pointed out the breakdown of the Hertzian contact model in nanoscale contact.[33] Recently, Wang et al. found that, when only the outmost surface step is involved, the circular flat punch model is more accurate to describe the contact with surface steps.[34] In this model, the load $F$ is linearly proportional to the indent depth $\delta$ as

$$F = 2E^* a \delta, \qquad (2)$$

where $a$ is the contact radius, and $E^*$ is the reduced modulus. Therefore, we utilize the flat punch model to fit the [110] compression behavior. In Fig. 2a, the fitted modulus is ~ 150.0 GPa, which is larger than the bulk modulus, 128.4 GPa, of bcc Nb in [110] direction.[35] The increase of elastic modulus has also been observed in other nanostructures, which is attributed to surface effects at nanoscale.[13][36] The predicted contact stress and contact area are also consistent with the simulated results in elastic stage, as shown in Fig. 2b.

Dislocation activities initiate at the moment the first atomic step is totally flatten when $\delta$ exceeds ~ 4.0 Å. Due to stress concentration, the fridges of surface step act as preferred sites for the nucleation of initial dislocation embryos. Fig. 3a depicts the activated <111>{110} slip systems under [110] compression. Each group of two neighboring {110} slip planes supply a V-shape pathway for dislocation gliding. After the top-most atom layer is flattened (Fig. 3b), two full dislocations with type of 1/2<111> nucleate beneath surface steps and glide toward the interior along the V-shape slip planes (Fig. 3c). These two dislocations intersect with each other subsequently and form a full dislocation of <001> type (Fig. 3d). Such a dislocation reaction can be expressed as

$$\frac{1}{2}[\bar{1}\bar{1}\bar{1}] + \frac{1}{2}[11\bar{1}] = [00\bar{1}]. \tag{3}$$

With further progress of compression, the full dislocation of <001> type dissociates into two new dislocations of type 1/2<111>, following the inverse reaction of Eq. (3). The new dislocations grow up and form two U-shape loops, and those dislocation loops then glide towards the central region of nanoparticle (Fig. 4a, b). In subsequent deformation, other U-shape dislocation loops directly nucleate from contact fridges as marked in Fig. 4c, which lead to the fluctuation of load in this stage. These dislocation loops glide and finally reach the free surface at the opposite side (Fig. 4d and e).

A jogged U-shape dislocation loop is highlighted in Fig. 4f ~ h. Its head segment is edge-like dislocation, and the lateral segments are screw dislocations. Because of the distinction in dislocation core structures,[23] the mobility of edge dislocations is higher than that of screw dislocations.[24][25] When a full dislocation loop begins to expand, its head segment moves faster than the lateral segments. After the head segment reaches and exhausts at free surface, the lateral screw segments are elongated and left behind, forming pure straight dislocations. These screw dislocations then gradually cross-slip towards surface. The activities of U-shape dislocation loop repeat in severe deformation stage, and many rod-like pure screw dislocations reside in the nanoparticle.

It is interesting to notice that the plastic deformation mechanisms in nanoparticle are clearly distinct from other bcc crystals. For bcc nanopillars under [110] compression, deformation twinning is the dominant mechanism indicated by both experiment and MD simulations.[37] For bcc bulk materials under [110] indentation, the plastic deformation proceeds mainly by the formation and de-pinning of prismatic dislocation loops, which further advance into interior.[38] For nanoparticles, dislocations with edge component are prone to exhaust at free surface rather than form prismatic loops, owing to the existing of lateral surfaces. The cross-slip of existing screw dislocations and new nucleation events sustain further plastic deformation.

**3.2 [111] compression**

The compression of bcc nanoparticle along [111] orientation is displayed in Fig. 5. Unlike the [110] compression, the compressive load increases smoothly following a power-law relation up to the yielding point ($\delta = \sim 9.0$ Å), as shown in Fig. 5 (a). The build-up of compressive load is then punctuated by a prominently sudden drop. Detailed information about the variations of contact stress and contact area is given in panel 5(b). Though the contact stress exhibits evidently serrated characteristics before yielding, it increases basically till to the yield point, then declines and remains at a low level. Instead, the contact area increases smoothly and steadily.

Compared with {110} planes, {111} families of crystal planes are not close-packed, and the crystal plane spacing is small. When compression is conducted along [111] direction, the loose-packed atoms in top layer are easily flattened and squeezed into the next atom layer. When yielding occurs, more than six atom layers have been flattened by the indenter. Accompanying the flattening of each layer, there is a drop of the contact stress. As pointed out by Wang et al.[34], once multiple surface steps have been flattened, the overall elastic compressive response of nanoparticles will approach the Hertzian contact model, which gives the relation of load $F$ and compression depth $\delta$ as

$$F = \frac{4}{3} E^* R^{1/2} \delta^{3/2}, \tag{4}$$

where $R$ is the radius of nanoparticle, and $E^*$ is the reduced modulus. We utilize Hertzian contact theory to fit the loading curve of [111] compression. In Fig. 5a, the fitted modulus in [111] direction is ~ 175.0 GPa, larger than 126.7 GPa of bulk Nb in [111] direction.[35] In Fig. 5b, based on the fitted modulus, we compare the predictions of Hertzian model and the simulated values for both averaged contact stress and contact area. Basically, Hertzian model can captures the elastic deformation in [111] compression. After yielding, loading curve evidently deviates from the elastic theoretical predictions.

Under [111] compressions, the stress concentration around the loose-packed surface facets is not as high as that of {110} surface facets. Only when multiple atom layers are flattened do dislocations appear. Initial twin embryos simultaneously nucleate in three equivalent <111>{112} slip systems (Fig. 6). Under further compression, twin embryos gradually grow up and form three twinning regions. Fig. 7 gives typical defect configurations at different compression depths. Because of the orientation of the activated slip systems, plastic deformation is highly localized in the twinning regions, which are gradually extruded from contact zone. While atoms in central region still maintain perfect lattice.

In order to identify the movement of dislocations on twin boundaries, we fetch a slice from the nanoparticle shown in Fig. 7e, which contains two neighboring {110} atom layers and cuts through two twinning regions. It is noted that one end of the twin boundary locates at the contact surface, and the other end terminates at the lateral free surface. During compression, partial dislocations nucleate at the intersection between the twin boundary and lateral surface, and then climb up along the {112} plane adjacent to the existing twin towards contact region. Under [111] compression, contact region serves as dislocation sinker, while the intersection between twin boundary and surface serves as dislocation source, and the gliding of the nucleated dislocation promotes twin migration.

These deformation features in nanoparticles are also unique compared to other bcc structures under [111] loading condition. For bulk bcc materials under [111] indentation, de-pinning and gliding of prismatic loops are still the typical behaviors.[38] For bcc nanopillars under [111] compression, deformation twinning is dominated, but only one twining slip system is activated.[37] When it comes to bcc nanoparticles, multiple twinning slip systems are activated, and deformation twinning simultaneously occurs on three slip systems owing to the weak lateral confinement.

**3.3 [001] compression**

Fig. 8 displays the loading curves of [001] compression. Under this compression, indenters firstly touch the (001) surface facets, and five (001) atom layers are flattened before yielding. Similar to [111] compression, Hertzian contact theory (Eq. 4) can characterize the elastic response, as shown in Fig. 8a. The fitted modulus is ~ 210.0 GPa, which is larger than that of the bulk modulus in [001] orientation, 133.2 GPa.[35] Yielding occurs at the strain of 0.108, and follows by an evident load drop. When the strain is larger than 0.13, the contact area begins to increase smoothly, and the flow contact stress stabilizes at ~ 8.8 GPa.

In this case, initial dislocation nucleates at some distance beneath the contact surface instead of the contact fringes, as shown in Fig. 9a. For [001] compression, there are four equivalent <111> slip directions, and both {112} and {110} planes are optional slip planes. When dislocations glide on {112} family of slip planes, both twinning and anti-twinning slip systems can be activated, such as $[11\bar{1}](\bar{2}11)$ and $[11\bar{1}](112)$. After nucleation, dislocation embryo grows up quickly and attaches the contact surface (Fig. 9b, c). One characteristic dislocation embryo is shown in Fig. 9e, and the corresponding activated slip systems are depicted in Fig. 9d. With the expanding of embryos, parts of initial dislocations cross slip onto twinning slip systems, meanwhile other parts cross slip onto anti-twinning slip systems or <111>{110} slip systems. Then, embryos on twinning slip systems evolve into deformation twinning regions, where a group of slipped perfect atoms are bounded by twin boundaries. While embryos on anti-twinning slip systems and {110} slip planes evolve into full dislocations. Owing to the gliding of 1/6<111> partial dislocations on twin plane, twinning regions are gradually broaden. Both deformation twinning and full dislocations coexist inside the deformed nanoparticle, as shown in Fig. 10. Unlike bcc nanopillars in which twin boundaries are flat planes,[37] twin boundaries here could not maintain planar characteristic, and their morphologies are greatly influenced by the heterogeneous deformation and stress distribution.

Under [001] indentation of bcc bulk materials, formation and gliding of prismatic dislocation loop remain the dominant role.[38] For [001] compression of bcc nanopillars, full dislocation gliding dominates without formation of prismatic dislocation loops.[39] However, in [100] compression of bcc nanoparticles, deformation twin and full dislocation coexist. The networks formed by full dislocations and twin boundaries play a key role for the stable flow contact stress.

## 3.4 Deformation mechanisms and dislocation density evolution

Three distinct plastic deformation modes in bcc nanoparticles are revealed. Different dominated mechanisms govern the evolution characters of dislocation densities, as shown in Fig 11a. And Fig. 11b gives the variation of fraction of defective atoms, i.e. the ratio of atoms constituting defects to the total atoms in nanoparticle, with respect to strain. Under [110] compression, full dislocation gliding is the main activity. After yielding, dislocation density stably reaches a maximum value, then decreases and fluctuates owing to the exhaustion of dislocation at free surface and new dislocation nucleation. Because of line defects dominating, fraction of defect atoms increases and fluctuates in low level. Under [111] compression, deformation twinning is the dominated mechanism, and full dislocation is rare, hence dislocation density is low. But the fraction of defective atoms keeps increasing after yielding due to expanding of twin boundaries. Under [001] compression, both full dislocation gliding and deformation coexist. After yielding, dislocation density keeps increasing in the whole plastic stage. This is because the entanglement between full dislocation and twin boundaries retards the exhaustion of dislocation at free surface. Since line and face defects coexist, the fraction of defective atoms maintains at a high level after yielding.

It should be pointed out that the loading with varying velocities from 0.01 Å/ps to 0.5 Å/ps have also been conducted. Our simulations reveal that the dislocation activities are quite similar, and the loading rate alters only the threshold of plasticity as expected.

## 4. Conclusion

This work provides orientation-dependent mechanical landscapes of bcc Nb nanoparticle by MD simulation. For elastic response, classical contact models should be carefully extended to nanoscale. The elastic behavior in [110] compression follows the flat punch contact model, while obeys the Hertzian contact model in [111] and [001] compressions. These features originate from the atomic morphology under indenter and the amount of atom layers flattened before yielding. In all cases, the elastic modulus of nanoparticle is larger than that of corresponding bulk material, which may be owing to the increasing significant surface effects at nanoscale. In bulk bcc materials, gliding of prismatic dislocation loop is the featured plastic behavior. However, depending on loading orientations, even one single crystal nanoparticle would exhibit non-trivial features associated with different mechanisms. Under [110] compression, full dislocation nucleating and gliding are the dominated activities, and dislocation density reaches a local maximum then decreases. Under [111] compression, deformation twinning is the primary mechanism, and the dislocation density maintains in a low level. And these two mechanisms coexist for [001] compression, in which dislocation density stably increases with compression. These results highlight the anisotropic behaviors of bcc nanoparticles and contribute directly to understand the underlying atomistic deformation mechanisms.

**Acknowledgement**:

Supports from the National Natural Science Foundation of China (Grant Nos. 11525209) are acknowledged. The calculations were performed on TianHe-1A at National Supercomputer Center in Tianjin.

**Figures:**

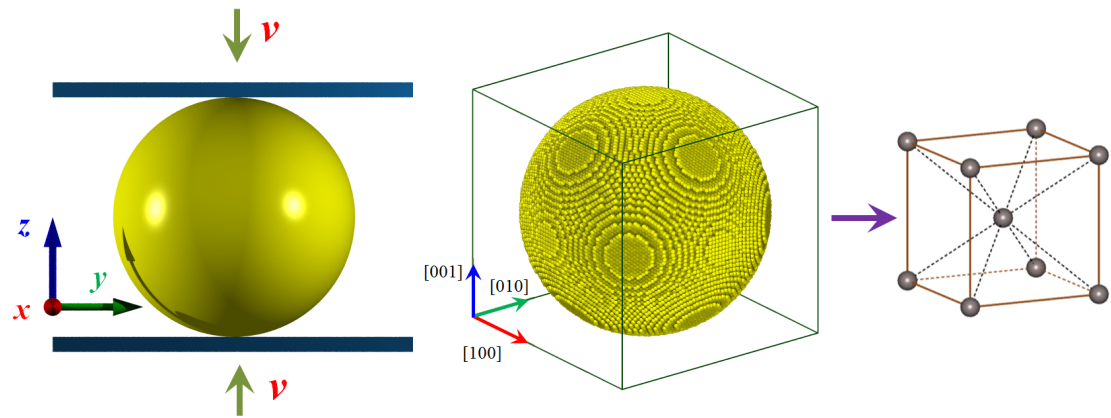

Fig. 1. Schematic of uniaxial compression of Nb nanoparticle and the initial configuration

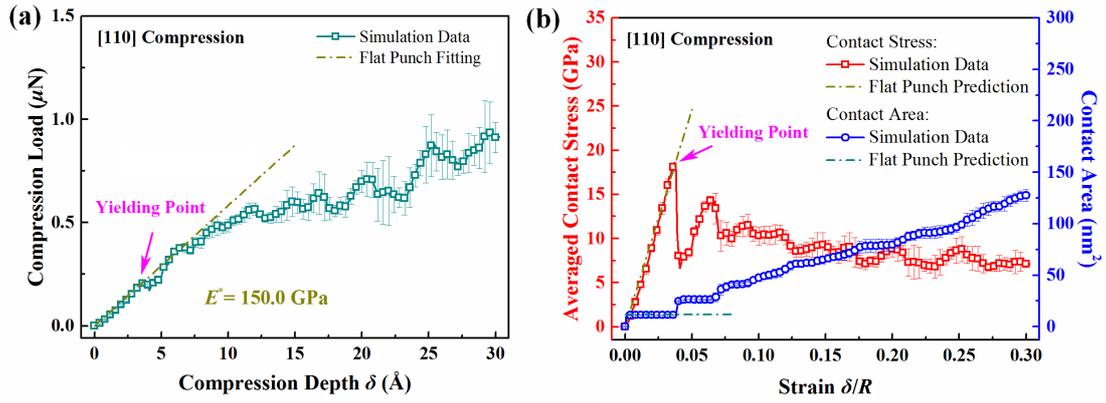

**Fig. 2**. (a) Loading curve and (b) variations of stress and contact area versus strain under [110] compression

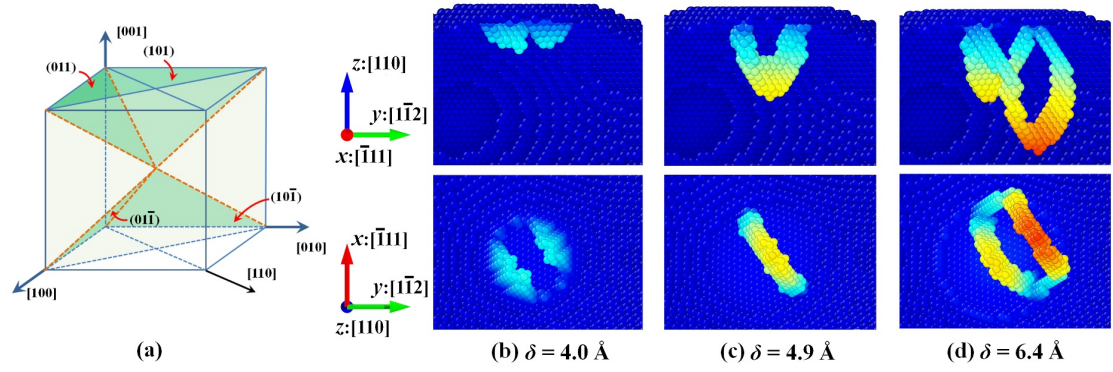

**Fig. 3.** (a) Activated <111>{110} slip systems for [110] compression, (b) ~ (d) nucleation and development of initial dislocations of [110] compression (Atoms in perfect bcc lattice are not shown for clearness, and atoms are colored according to their position relative to the center of nanoparticle. Following atomic figures use the same atom coloring method)

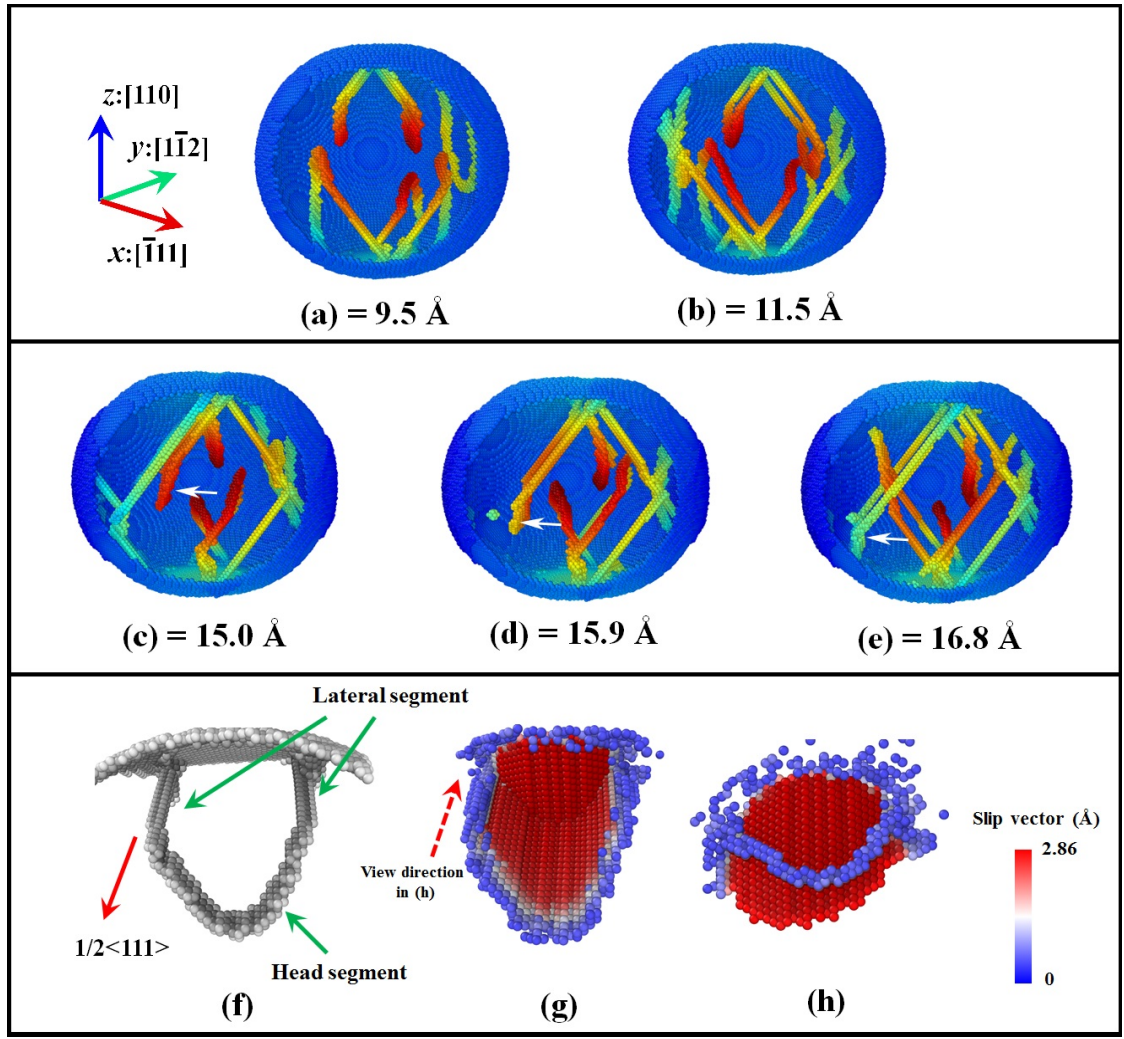

**Fig. 4**. (a) ~ (e) Full dislocation evolution inside nanoparticles, (f) a highlighted single full dislocation loop and its different views in (g) and (h)

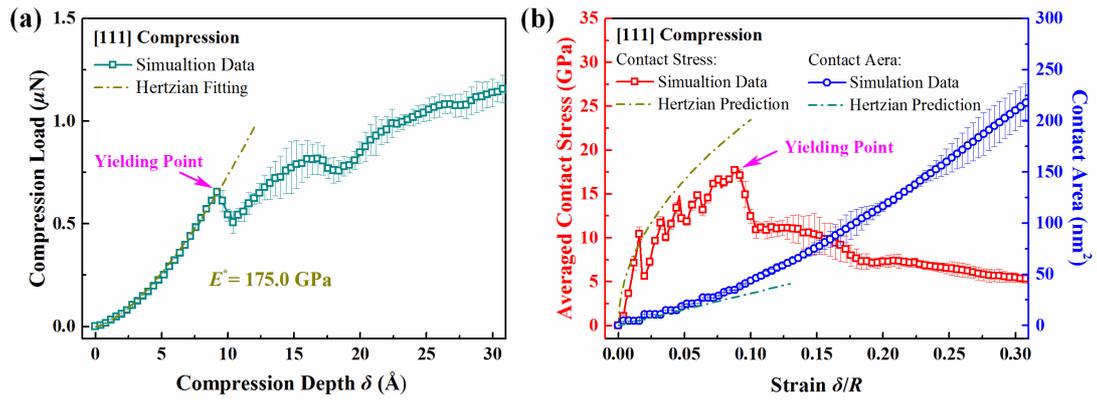

**Fig. 5**. (a) Loading curve and (b) variations of stress and contact area versus strain under [111] compression

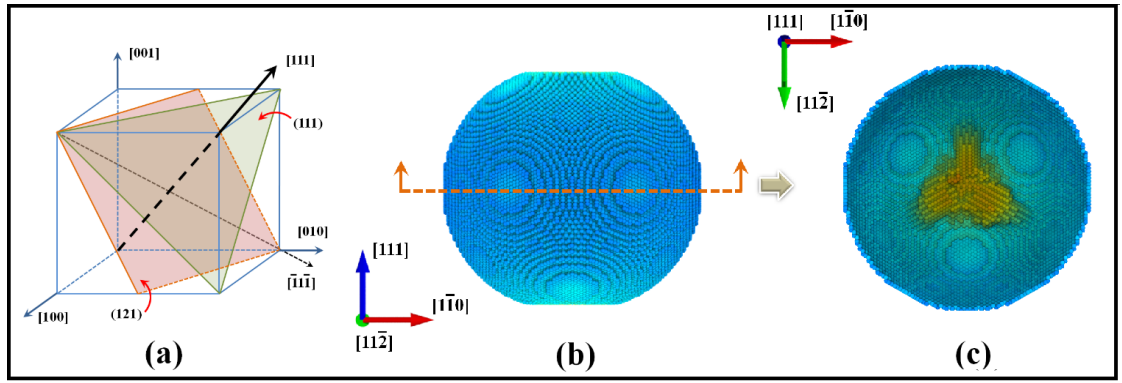

**Fig. 6**. (a) Activated <111>{112} slip systems for [111] compression, (b) front view and (c) bottom view of initial twin embryos under the indent

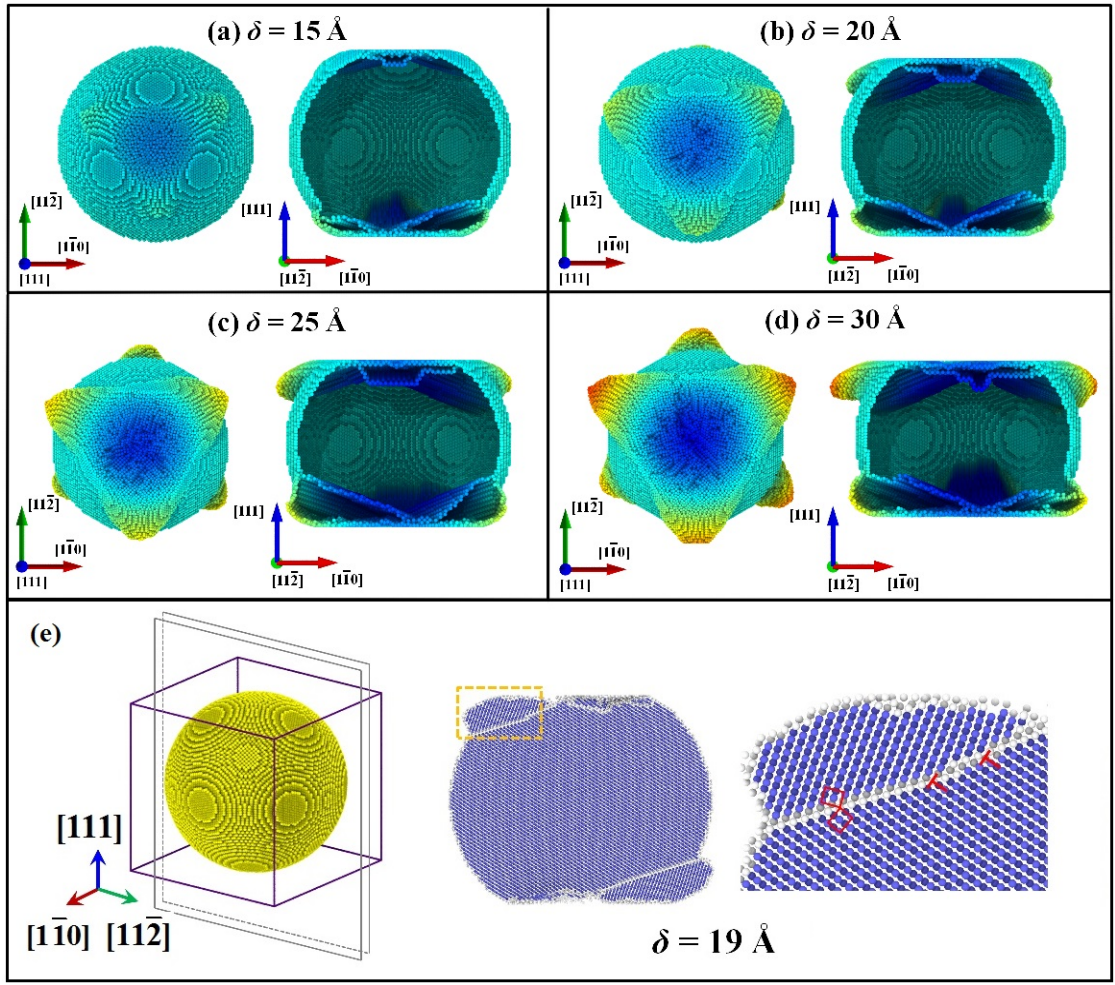

Fig. 7. (a) ~ (d) Development of twinning regions under indenter for [111] compression, (e) twin boundary migration for [111] compression

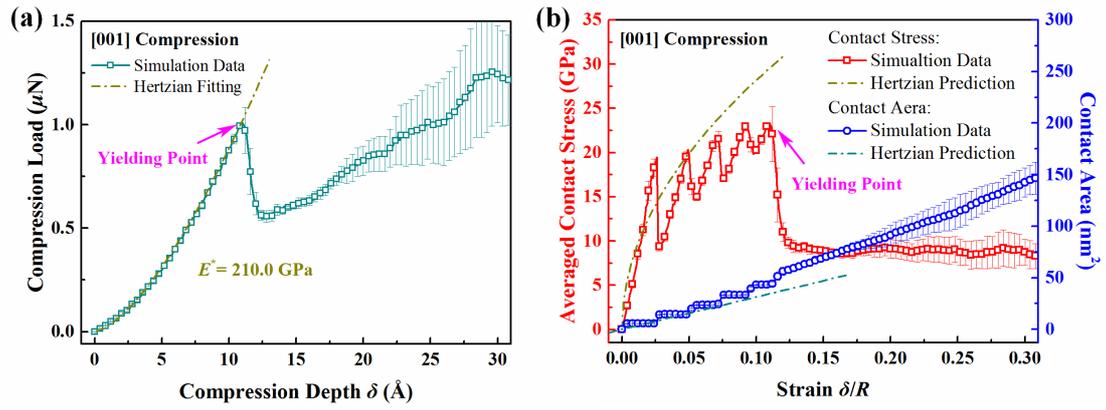

**Fig. 8**. (a) Loading curve and (b) variations of stress and contact area versus strain under [001] compression

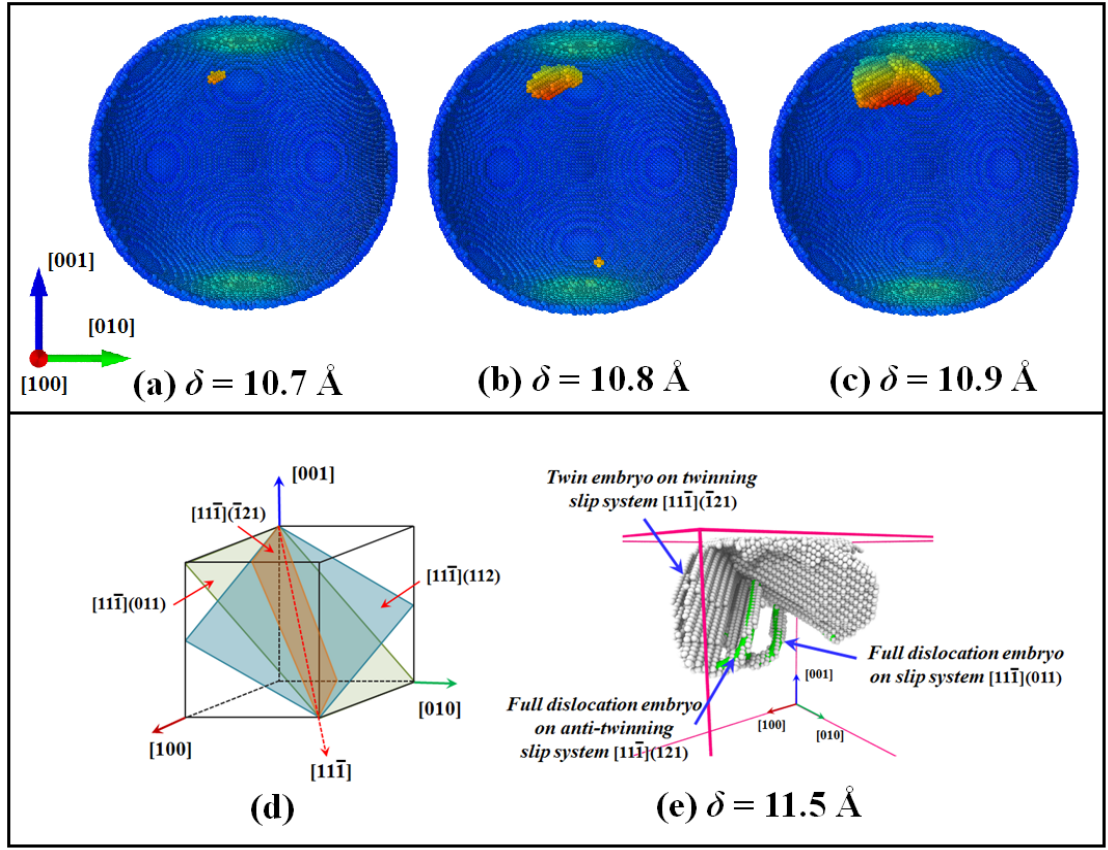

**Fig. 9.** Evolution of the homogeneously nucleated dislocation embryos under surface for [001] compression (a) ~ (c), (d) and (e) show dislocation embryos on different slip systems

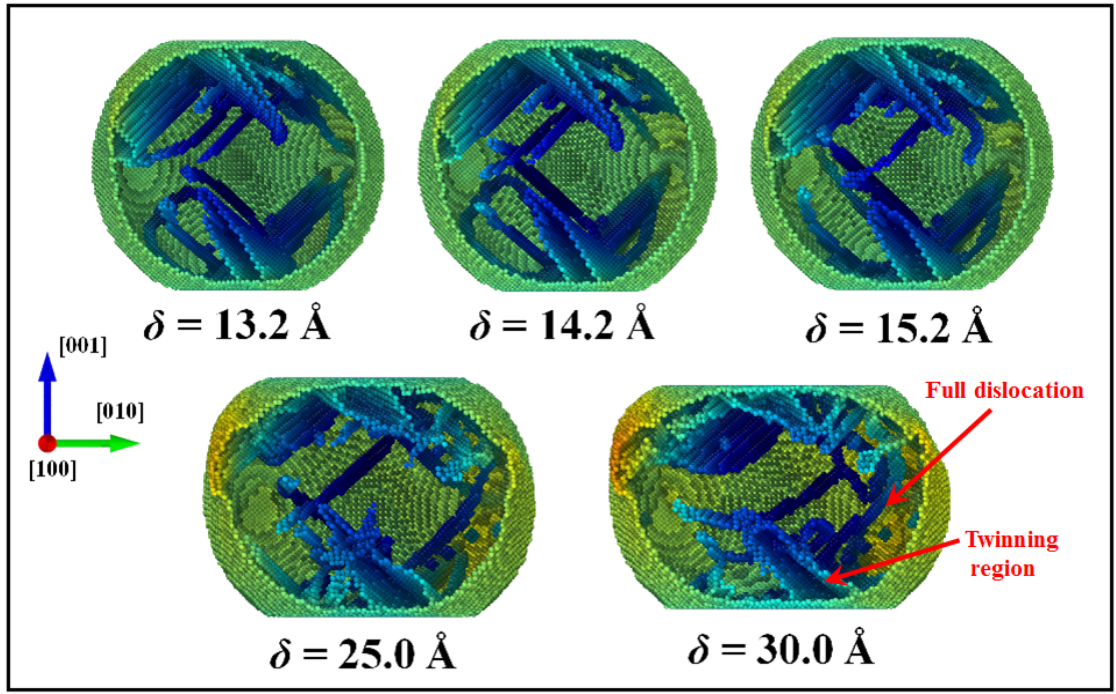

**Fig. 10**. Dislocation and twin evolution coexist inside the deformed nanoparticle

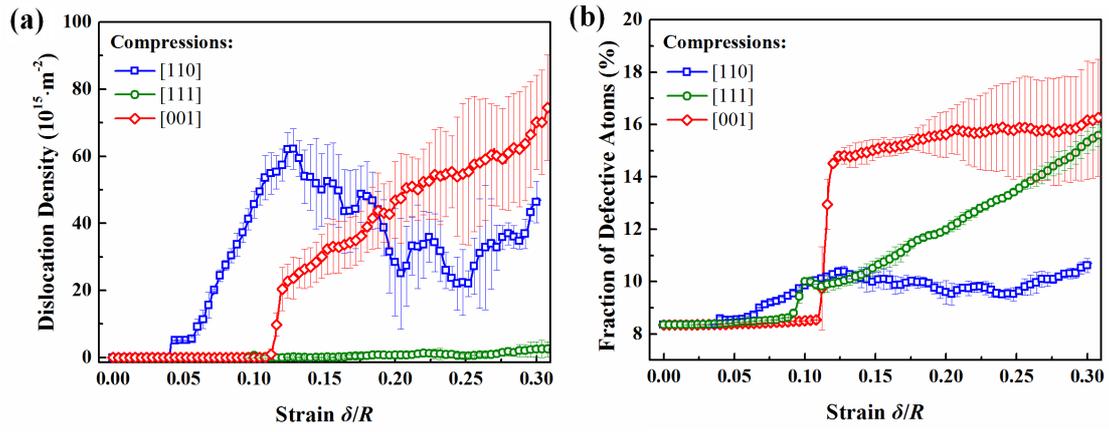

**Fig. 11**. (a) Evolution of dislocation density and (b) evolution of fraction of defective atoms during compression of nanoparticles